\documentstyle[12pt,aasms4]{article}
 
\slugcomment{To appear in the Astrophysical Journal (letters)
(February 20, 1998 issue)} 
 
\begin{document}
 
\title{Erratum for: ~An X-ray Cluster at  Redshift 2.156?}
 
\author{C.L. Carilli}
\affil{National Radio Astronomy Observatory, P.O. Box O, Socorro, NM,
87801 \\} 
\authoremail{ccarilli@nrao.edu}
\author{D.E. Harris}
\affil{Smithsonian Astronomical Observatory, 60 Garden St., Cambridge,
MA, 02138 \\} 
\author{L. Pentericci, H.J.A. R\"ottgering, G.K. Miley, and M.N. Bremer}
\affil{Leiden Observatory, Postbus 9513, 2300 RA Leiden, The
Netherlands \\}  
 
\vskip 0.3in
\centerline{\bf ERRATUM}

A third possible explanation for the X-ray emission from the z = 2.156
radio galaxy 1138$-$262 is thermal emission from a very dense,
sub-cluster `halo', perhaps associated with the (forming) cD galaxy on
a scale $\le$ 100 kpc. For  instance, increasing the gas density from
0.01 cm$^{-3}$ to 0.1 cm$^{-3}$  would decrease the required hot gas
mass by the same factor, and could possibly alleviate constraints on
cosmological structure formation models. The pressure in this gas
would be very high (10$^{-9}$ dynes cm$^{-2}$), comparable to the
pressure in the optical line emitting  
nebulosity and to the minimum pressure in the radio source, and the
cooling time would be short ($\le$ few$\times$10$^{8}$ years).
Circumstantial evidence for such very dense, hot gas enveloping some
high z radio sources has been reviewed by Fabian et al. (1986). Fabian
(1991) suggests that in some cases the `inferred pressures are close
to the maximum that can be obtained by gas cooling in a potential well
of a galaxy' (ie. the cooling time $\approx$ gravitational free-fall
time), and he designates such systems as  `maximal cooling
flows', with implied cooling flow rates up to 2000 M$_\odot$
year$^{-1}$.  High resolution  X-ray imaging with AXAF should be able
to test whether 1138$-$262  has a normal cluster atmosphere, a `maximal
cooling flow', or an unusually X-ray loud AGN. 

\clearpage
\newpage

\centerline{\bf References} 

Fabian, A.C., Arnaud, K.A., Nulsen, P.E.J., and Mushotzky, R.F. 1986,
Ap.J., 305, 9

Fabian, A.C. 1991, in `Clusters and Super Clusters of
Galaxies,' ed. A.C. Fabian (Dordrecht: Kluwer), p. 151.

Acknowledgments: We would like to thank W. Forman and F. Owen
for useful discussions on this issue.

\clearpage
\newpage
\end{document}